\documentclass[fleqn,usenatbib]{mnras}

\usepackage{color}

\usepackage[T1]{fontenc}

\DeclareRobustCommand{\VAN}[3]{#2}
\let\VANthebibliography\thebibliography
\def\thebibliography{\DeclareRobustCommand{\VAN}[3]{##3}\VANthebibliography}

\usepackage{graphicx}	
\usepackage{amsmath}	
\usepackage{amssymb}	
\usepackage{tabularx} 
\usepackage{gensymb} 

\def\hcMpc{$h^{-1}\rm{cMpc\ }$}
\def\hcMpcvol{$h^{-3}\rm{cMpc^3\ }$}
\def\Lya{Ly-$\rm{\alpha}\ $}
\newcommand{\code}[1]{\textsc{\small #1}}

\usepackage{newtxtext,newtxmath}

\title[Large scale Cosmic Dawn]{21cm signal predictions at Cosmic Dawn and Reionization with coupled radiative-hydrodynamics }

\author[N J. F. Gillet et al.]{
N. J. F. Gillet$^{1}$\thanks{E-mail: nicolas.gillet@astro.u-strasbg.fr },
D. Aubert$^{1}$, 
F. G. Mertens$^{2,3}$, 
and P. Ocvirk$^{1}$
\\
$^{1}$Observatoire Astronomique de Strasbourg, Université de Strasbourg, CNRS UMR 7550, 11 rue de l'Université, 67000 Strasbourg, France\\
$^{2}$Kapteyn Astronomical Institute, University of Groningen, PO Box 800, NL-9700 AV Groningen, the Netherlands\\
$^{3}$LERMA, Observatoire de Paris, PSL Research University, CNRS, Sorbonne Universite, F-75014 Paris, France
}

\date{Accepted XXX. Received YYY; in original form ZZZ}

\pubyear{2020}

\begin{document}
\label{firstpage}
\pagerange{\pageref{firstpage}--\pageref{lastpage}}
\maketitle

\begin{abstract}
The process of heating and reionization of the Universe at high redshift links small scale structures/galaxy formation and large scale inter-galactic medium properties.  Even if the first one is difficult to observe, an observation window is opening on the second one, with the promising development of current and future radio telescopes.   They will permit to observe the 21cm brightness temperature global signal and fluctuations.  The need of large scale simulations is therefore strong to understand the properties of the IGM that will be observed.  But at the same time the urge to resolve the structures responsible of those process is important.   We introduce in this study, a coupled hydro-radiative transfer simulations of the Cosmic Dawn and Reionization with a simple sub-grid star formation process developed and calibrated on the state of the art simulation CoDaII.  This scheme permits to follow consistently dark matter, hydrodynamics and radiative transfer evolution's on large scales, while the sub-grid models bridges to the galaxy formation scale.  We process the simulation to produce 21cm signal as close as possible to the observations. 
\end{abstract}



\section{Introduction}
\label{Sec:Introduction}


Despite the fact that the first billion years are full of events, the properties of the Universe between the emission of the Cosmic micro-wave background (CMB) and redshift z=6 are still poorly constrained by observations.  It sees the Cosmic Dawn (CD), the birth and growth of the first structures, stars and galaxies, as well as the cosmological change of properties of the inter-galactic medium (IGM), from cold and neutral to hot and ionized during the Epoch of Heating (EoH) and Reionization (EoR).  Observational prospects seems promising, with e.g. high redshift galaxies probed by the James Webb Spatial Telescope (JWST) or the avalanche of data from the current and in-development radio telescopes that will measure the IGM properties on large scales.  In this work, a special focus will be put on this latter type of instruments such as: \textit{LOw Frequency ARray} (LOFAR; \citealt{LOFAR2013}), which acquires data between 200MHz and 110 MHz (between redshift 6 to 12), therefore focusing on the end of the EoH and the EoR.  The LOFAR EoR Key Science Project has recently put upper limits on the power spectrum of the cosmic 21cm signal at redshift 9.1 \citep{Mertens2020}.  Another instrument, the \textit{New Extension in Nan\c{c}ay Upgrading loFAR} (NENUFAR; \citealt{NENUFAR2012}), acquiring data between 85MHz and 30MHz (from redshift 16 to 45) and therefore overlapping with the frequency range of claimed detection of the global signal at redshift 17 of the EDGES instrument \citep{Bowman2018}.  In parallel, theoretical modeling of the physics and the signal is ongoing and aims at following the structure and galaxy formation on small scales and its impact on the properties of the IGM on large scales.   The theoretical challenge is to do both : resolving the birth and properties of the first galaxies, their photons emission (Lyman-$\alpha$, x-rays, UV for example) and tracking the evolution of the IGM properties on cosmological distances. 

Several groups address this challenge by using analytical models of galaxy formation (most often base on the local collapse mass fraction or halos mass function) and  semi-numerical treatments of the Reionization \citep[e.g.][]{Visbal2012,  Fialkov2014, 21cmfast}.  These methods have the advantage to have a comprehensible set of galaxy formation parameters and to be computationally efficient.  Alternatively, others push to directly solve all scales with coupled hydrodynamics-radiative transfer simulations \citep[e.g.][]{Gnedin2016, Semelin2017, CoDaI, CoDaII}.  But the trade-off between resolution and volume makes those simulations difficult to realize and costly, while being still limited in the range of halo masses or cosmological scales that can be probed. A final alternative is to perform the radiative transfer in post-process on top of dynamics-only simulations.  It uses high resolution dark matter halos to support a galaxy formation model while providing a realistic propagation of photons \citep[e.g.][]{Chardin2017, Kulkarni2019, Ross2019}.  The gain in computational time can be significant compared to fully coupled simulation but this method cannot probe the full extent of the respective feedbacks of matter and radiation.

In this article, we present an other alternative to produce large scale-simulations ($>250$ cMpc needed for proper IGM properties \citealt{Iliev2014, Kaur2020}) in the context of current and future radio experiments of the Cosmic Dawn.  It relies on fully-coupled radiative transfer-hydrodynamics and a sub-grid model of galaxy formation at the necessarily moderate resolution (1 cMpc) on such volumes.  One of the most pressing challenges is the lack of source formation during the Cosmic Dawn due to the limited resolution in large simulated volumes.  Standard sub-grid star formation models cannot create sources in an efficient manner and alternatives must be developed, such as the one we describe here.  We propose that the unresolved star formation could be based on state-of-the-art high resolution simulations of the EoR such as CoDaII \citep{CoDaII}.  This technique is implemented in the EMMA cosmological simulation code \citep{EMMA} and is demonstrated in the following sections.  It permits to have a fully coupled evolution of the radiative field and the IGM gas while the sub-grid source model takes care of the non-resolved structure formation and evolution.  We show that this methodology leads to viable and consistent predictions of the 21 cm radio signal from the Cosmic Dawn. We introduce the calibrated sub-grid source formation model in Section \ref{Sec:Methodology} and then discuss the resulting large-scale 21cm signal predictions in Section \ref{Sec:21cm signal}.


\section{Source formation model and simulation}
\label{Sec:Methodology}


In this study we use the 'full-physics' cosmological simulation code for reionization EMMA \citep{EMMA} in a large scale/low resolution mode.  We extend the code with a new empirical source (star/galaxy) formation model based on the CoDaII simulation that provides more flexibility at high redshift than standard methods and we also add simple prescriptions for the prediction of the 21cm signal.  


\subsection{Sources}
\label{Sec:Sources}


The challenge of large-scale/low-resolution simulations is to assign a production of ionizing photons per volume unit despite the lack of dense, non-linear structures in simulations.  At high redshift ($z>6$) the main sources of UV photons are young massive stars~:  we have to find a way to assign to each resolution element a star formation rate (SFR), to follow the creation of stars, i.e. the sources of UV photons.  A sub-grid model has to be constructed, that would assign a production of photons as function of the local structure formation.  

Classically, semi-analytical galaxy formation models rely on the underlying dark matter collapse fraction and halos mass function (for examples \citealt{Fialkov2014, 21cmfast} ).  It assumes that galaxies form in halos and the star formation depends on the halos mass.  In this study we take an alternative approach. Instead of trying to resolve and follow dark matter structures formation and evolution, we simply suppose that a fraction of the gas will be star forming on mega-parsecs scales.  We develop an empirical large-scale galaxy formation model based on the results of the state of the art high-resolution, hydro-radiative simulation of the Reionization, CoDaII \citep{CoDaII}.

\subsubsection{Star formation in CoDaII}
\label{Sec:Star formation in CoDaII}

The CoDaII simulation has a box of 64\hcMpc side sampled on a Cartesian grid of $4096^3$ elements.  In a very standard manner, the production of stellar particles during the simulation is driven by a SFR density computed at each time step, according to:
\begin{equation}
         SFR_{\varphi}^H \propto \epsilon_* \rho^{1.5}, \rm{where} \ \rho>\rho_*.
         \label{Eq:SFR_CoDaII}
\end{equation}
In the CoDaII simulation, the SFR is directly proportional to the density at power 1.5, where $\epsilon_*=0.42$ is the star formation efficiency and $\rm{\rho_*/f_{\Omega}=\Delta_*=50 }$ is the star formation density threshold ($\rm{ f_{\Omega} = \Omega_b / \Omega_m }$ is the baryonic fraction). 

In post-processing, we degrade the simulation outputs on a coarse grid of $64^3$ cells corresponding to our 1\hcMpc goal resolution.  In each cells we compute a 'coarsened' density contrast ($\rm{\Delta=\rho/\langle \rho \rangle}$) and a 'coarsened' star formation rate, using 10 snapshots between redshift 5.7 to 15. This post-processed SFR density of CoDaII simulation is computed in each coarse 1\hcMpc cell as the sum of the stellar particle masses younger that 10 Myr, divided by 10 Myr. Hereafter, physical scale quantities are annotated with the letter '$\varphi$' and comoving one with a 'c'. Low and high resolution quantities are annotated with 'L' and 'H', respectively and refer to 1\hcMpc or to the original CoDaII resolution (15.625$h^{-1}\rm{ckpc}$).  
One might directly apply Eq \ref{Eq:SFR_CoDaII} on the low resolution grid, but it results in a too permissive star formation at high redshift (z$\sim$30).  Indeed, the density contrast are smaller at low resolution at every redshift, therefor having one fixed density threshold would produce an almost flat cosmic SFR evolution with redshift, too much star at high redshift or not enough at low redshift.  We need to change the threshold parameter to mimic the sub-grid collapse structures and control the SFR at high and low redshifts.  Furthermore, the classical scheme applied at low resolution cannot take into account the sub-grid quenching of the Reionization on the smallest galaxies.  To take this effect into account we derive an empirical model based on the outputs of the CoDaII simulation.

\subsubsection{Sub-grid star formation rate}
\label{Sec:Sub-grid star formation rate}

\begin{figure*}
\includegraphics[width=1.\textwidth]{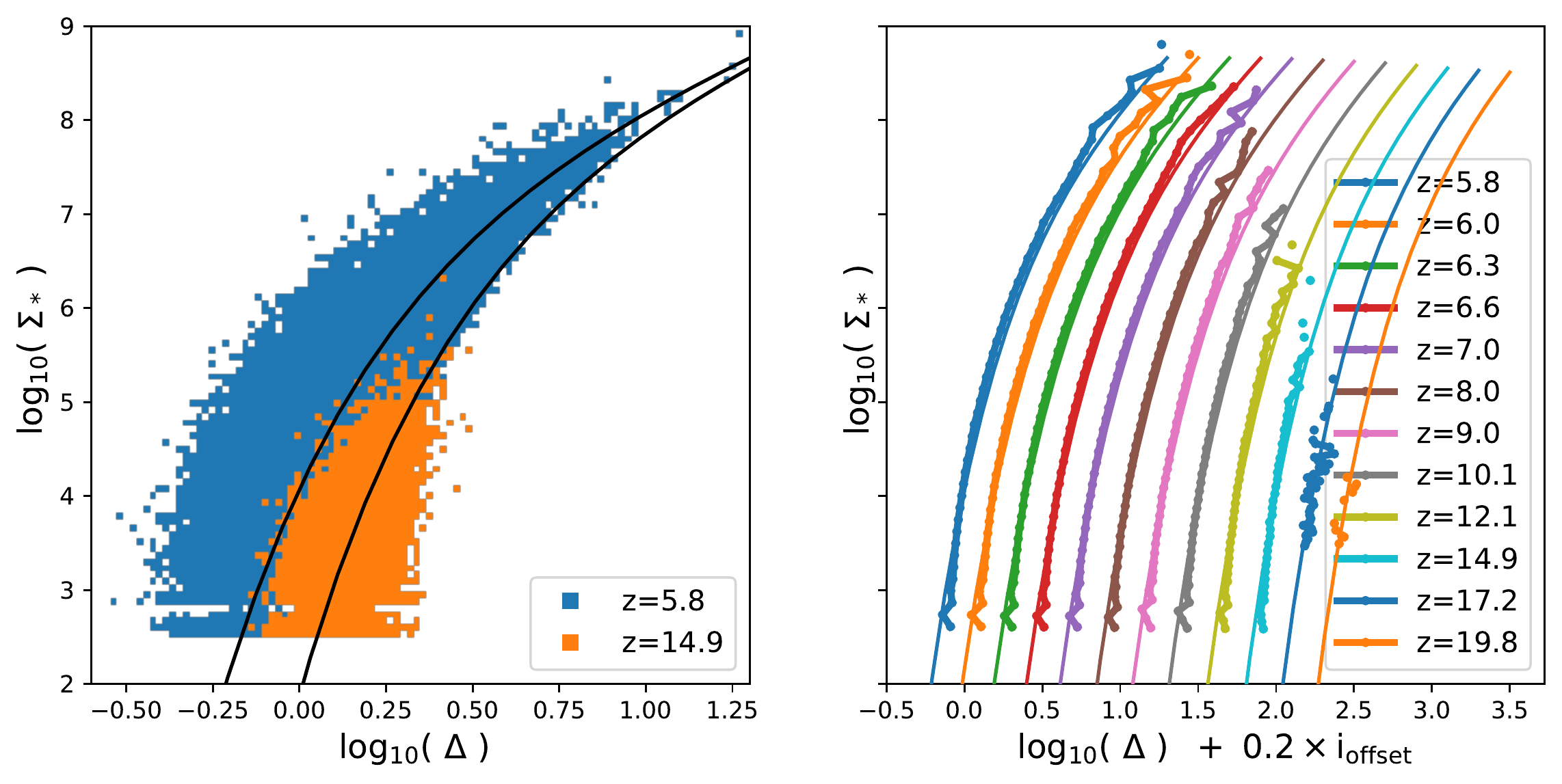}
\caption{ The proxy to the star forming gas density ($\rm{\Sigma_*}$) as function of density ($\rm{\Delta}$): On the left, the distribution of all coarse cells are shown at redshift 6 and 15, with the fitted function in black.  The right panel presents the mean relations and their fits at all available redshifts. Each redshift relation is shifted by 0.2 dex for clarity.  Note that at redshift 17 and 20 we do not have access to the full resolution data, but to the $2048
^3$ cubes, which explains the difference in resolution accessible in $\rm{\Sigma_*}$ for those two redshifts. }
\label{Fig:Sstar_vs_Delta}
\end{figure*}

In the CoDaII case, each coarse cell (1\hcMpc) is composed of $64^3$ high resolution cells.  Each of them can be star forming ($SFR\propto \Delta^{1.5}$, c.f. Eq \ref{Eq:SFR_CoDaII}).  Therefore we derive that the low resolution SFR on co-moving scale is defined as follow:

\begin{equation}
         SFR_{c}^L = \bar{\epsilon_*} \Sigma_*^L  \frac{1}{\rm{a_{exp}^{1.5}}},
    \label{Eq:SFR_model}
\end{equation}

where $\bar{\epsilon_*}$ is the proportionality factor that absorb all the constants.  The expansion factor dependence comes from the physical to comoving transformation and the power 1.5 dependence to the density (c.f. Eq \ref{Eq:SFR_CoDaII}).  And we define $\Sigma_*$ as the star forming gas density at the power 1.5 in each coarse cell: we call it the 'proxy to the star forming gas: PSFG'. It is computed in the CoDaII simulation post-processed outputs as the sum of the density at power 1.5 of star forming cells:

\begin{equation}
     \Sigma_*^L = \sum {(\Delta_{i}^{H})}^{1.5}\ \rm{where\ } \Delta>\Delta*
\end{equation}

where the iterator \textit{i} stands for each of the $64^3$ high resolution cells in a coarse cell of 1\hcMpcvol.  Fig. \ref{Fig:Sstar_vs_Delta} presents the PSFG for all coarse cells of 1\hcMpcvol as function of the over-density.  The left panel presents the distribution of ($\Delta^L,\Sigma_*^L)$ pairs in CoDaII at redshift 6 and 15.  At high density, $\Sigma_*$ follows a power law as a function of the density contrast with a unit slope.  The PSFG decreases sharply as the density contrast becomes smaller.  And the scatter around the overall trend is large (for example, at $\Delta=1$, $\Sigma_*$ covers almost 4 orders of magnitude at redshift 6).  The dispersion increases as the density decreases, and at the same time a hard minimum is set, imposed by the CoDaII simulation parameters ($\rm{} \Sigma_{*,min}=50^{1.5}$, corresponding to a single high resolution cell above the star formation threshold in one coarse cell).  

For the sake of simplicity, we model the mean behavior of the ($\Delta^L,\Sigma_*^L$) relation.  $\Sigma_*$ behave as a power law with respect to the density with an exponential cutoff at the low-density end.  This model is purely empirical and does not take into account the dispersion induced by the variance in structure formation.  But it does take into account of the underlying stellar and radiative feedback on the gas density implemented in the CoDaII simulation. The density of star forming gas is parametrized as follow:  

\begin{equation}
    \Sigma_* =  \epsilon_{\Sigma_*} \Delta 10^{-\Delta_{\Sigma_*}/\Delta },
\label{Eq:Sigma_star}
\end{equation}

where $\epsilon_{\Sigma_*}$ is fitted at redshift 6 and kept constant at all higher redshifts ($\rm{log_{10}}(\epsilon_{\Sigma_*})=7.55$). 
Then, $\Delta_{\Sigma_*}$ is adjusted at each redshift independently.  

The evolution of the parameter $\Delta_{\Sigma_*}$ with redshift is obtained here for the CoDaII simulation.  The mean evolution with redshift of the PSFG as function is shown of the right panel of Fig \ref{Fig:Sstar_vs_Delta}.  Nevertheless, its evolution can be freely parametrized (empirically or physically) to explore or accommodate different scenarios and models of star formation, for example the inclusion of POPIII stars, or more simply modulate the time evolution of the cosmic SFR.  For sake of simplicity we consider a linear evolution of $\Delta_{\Sigma_*}$ with redshift, which is roughly consistent with the evolution given by CoDaII.  

\subsubsection{Star formation space distribution}
\label{Sec:Star formation space distribution}

\begin{figure*}
\includegraphics[width=.33\textwidth]{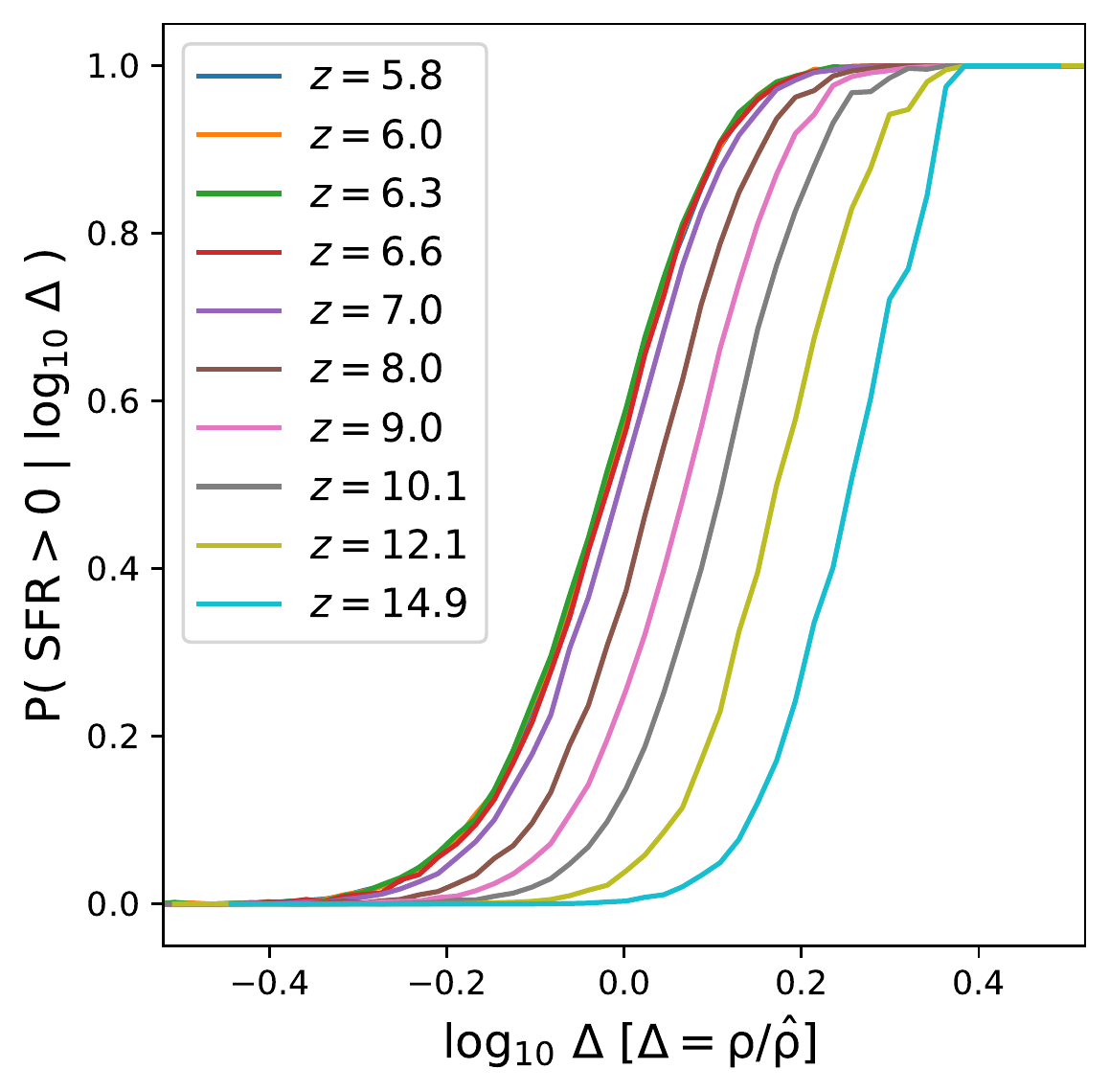}
\includegraphics[width=.33\textwidth]{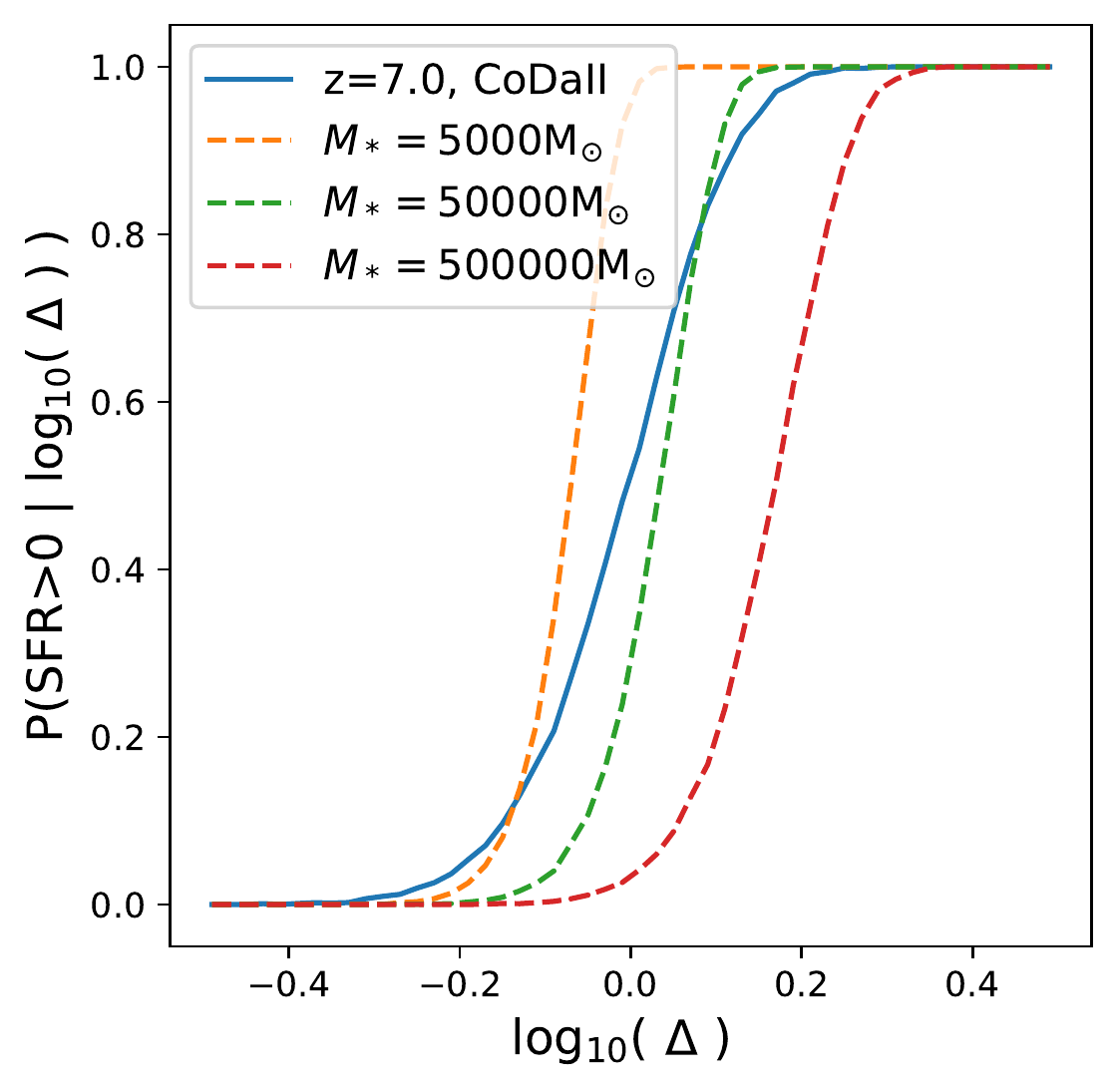}
\includegraphics[width=.33\textwidth]{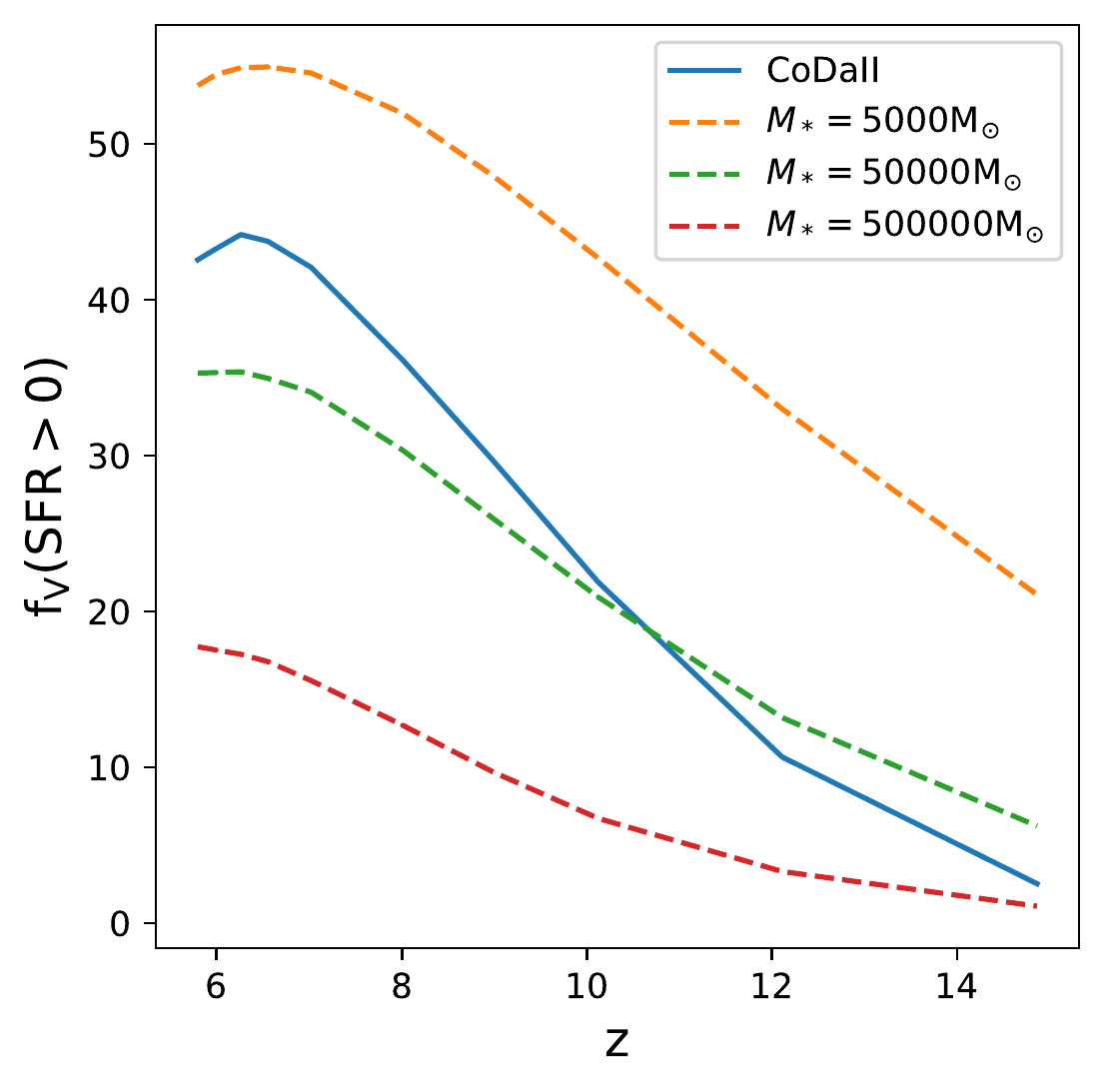}
\caption{ On the left panel, the probability for a cell of $1h^{-3}.\rm{cMpc^3}$ of the CoDaII simulation to have a non-zero SFR as function of density and redshift.  Line from redshift 5.8 to 6.6 are almost identical and stack.  
On the middle panel, the same as the left, but only at redshift 7. The CoDaII simulation is the blue line, while the dash orange, green and red are the sub-grid model with $M_*=5000, \ 50000, \ 500000 \ M_{\odot}$ respectively.  
On the right panel the volume filing factor of non-zeros SFR cells of $\rm{1h^{-3}.cMpc^3}$, for the CoDaII in blue and the sub-grid models are the same as in the middle panel.  }
\label{Fig:PSFRsup0}
\end{figure*}

At this stage every cell has a non-zero SFR.  But, as we expect to have more star formation in the densest regions, we also expect to have no star formation in the most under-dense ones and in between a certain stochasticity.  The left panel of Fig. \ref{Fig:PSFRsup0} illustrates the stochasticity by presenting the probability for a cell of 1\hcMpcvol to have a non-zero SFR, as function of the density contrast and redshift in the CoDaII simulation.  The transition is smooth between high densities that always form stars and the low-density regions that do not.  And this transition evolve with redshift, shifting toward low-density regions with time.  At $z=6$, a 1\hcMpcvol volume with an average density has a 50\% probability to be star-forming.  Another way to visualize the stochasticity and the spatial distribution of the star forming region is to look at the volume filling factor of star forming cells, presented on the right panel of Fig. \ref{Fig:PSFRsup0}.  The blue line shows the SF volume filling factor of the CoDaII simulation, coarsened on scales of 1\hcMpc.  The fraction of volume that form stars rise with redshift, with a maximum just below 50\% between redshift 6 and 7.  It means that, in the CoDaII simulation, almost half of the volume of the Universe is star-forming at redshift 7, smoothed on scale of 1\hcMpc.

The local variations introduce above and the resulting SF spacial distribution will set the spatial evolution of the reionization process.  It will affect the HII bubble size distribution and evolution, and   the 21cm temperature brightness power spectrum (PS) too.  Therefore we introduce here one way to control the star formation distribution in our simulations.  We use a minimum stellar mass $M_*$ and the star formation process is discretized in stellar particles.  With the same scheme as in CoDaII, the number of stellar particle created is drawn from a Poisson distribution.  The mean SFR of a coarse cell is set by Eq. \ref{Eq:SFR_model}.  Then the mean stellar mass is obtain by multiplying by the time step ($dt$), and the mean number of stellar particles is therefor obtain by dividing by the stellar mass particle $\bar{N_*} = SFR^L_c \times dt / M_*$.  In the end, the parameter $M_*$ does the same as in high-resolution runs. It permits to set a minimum SFR in a cell and to cut star formation where it is it too low.  However, the physical meaning of $M_*$ is different.  Here it encompass the local variations due to the star formation and unresolved structure formation at the same time.  We apply our new parametrization of the source formation on the outputs of the CoDaII simulation.  The impact of $M_*$ on the star formation process is illustrated on the middle and right panel of Fig. \ref{Fig:PSFRsup0} with different $M_*$: $5.10^3\rm{M_{\odot}}$, $5.10^4\rm{M_{\odot}}$, $5.10^5\rm{M_{\odot}}$ (orange, green and red respectively).  
This parameters controls the distribution of the star formation as a function of the density, as shown on the middle panel of Fig. \ref{Fig:PSFRsup0}, which automatically translates to the volume filling factor, shown in the right panel.  Interestingly, as shown after, as the cosmic star formation density is mostly set by the heaviest regions, these parameters does not affect the global SFR.  Therefore, the global SFR and its spatial distribution are almost independent with this parametrization.  The parameter $M_*$ permits to choose between a "diffuse" or a "biased" SFR distribution.


\subsection{Simulation's set}
\label{Sec:Simulation's set}


The previously presented star-formation model and an on-the-fly computation of the 21cm signal (presented hereafter) have been added in the hydrodynamics-radiative transfer code EMMA \citep{EMMA}.  It permits to realize cosmological simulations of the of the CD, EoH and EoR by coupling the evolution of dark matter, baryonic matter, source formation and radiative transfer. 
\subsubsection{Specifications}
\label{SEC:Specifications}

We produce a $(512 h^{-1}\rm{cMpc})^3$ simulation with a resolution of 1\hcMpcvol.  The simulation's specifications are listed on Tab. \ref{Tab:specs}.  The source formation starts at redshift 30 and the actual speed of light is used for the radiative transfer, to avoid artifacts as reported in \citep{Deparis2019,Ocvirk2019}. X-rays are included in those simulation and it is important to recall that \Lya radiation is not include in the simulation yet.  The study of X-ray and \Lya will be done in the follow-up study.

\begin{table}
\begin{tabularx}{\columnwidth}{ >{\raggedright\arraybackslash} X  >{\raggedleft\arraybackslash} X  }
 \hline
 \multicolumn{2}{c}{Cosmology (Planck 18)} \\
 \hline
 $\rm{\Omega_{\Lambda}}$ & 0.6889 \\
 $\rm{\Omega_{m}}$  & 0.3111    \\
 $\rm{\Omega_{b}}$ & 0.04897 \\
 h & 0.6766 \\
 $\rm{\sigma_8}$ & 0.8102 \\
 $\rm{n_{spec}}$ & 0.9665 \\
 
\hline
\multicolumn{2}{c}{Stars} \\
\hline
${\rm{log_{10}}}\ \overline{\epsilon_{*}}$ & -7 \\
${\rm{log_{10}}}\ {\epsilon_{\Sigma_*}}$ & 7.55 \\
$M_*$ & $ 10^6\ \rm{[M_{\odot}]}$  \\
$t_*$ & 10 $\rm [Myr]$  \\
$z_{ON}$ & 30 \\

\hline
\multicolumn{2}{c}{Radiation} \\
\hline
Stellar ionizing emissivity & 4.32  $\rm 10^{46} [ph.s^{-1}.M_{\odot}^{-1}]$ \\
$f_{esc}$ & 0.05 \\
$\rm \langle E_{UV} \rangle$ & 20.65  $\rm [eV]$ \\
$\rm \langle E_{X-ray} \rangle$ & 224.56  $\rm [eV]$ \\
$\rm \sigma_{UV}$ & 2.381 $\rm \times 10^{-22}\ [m^2]$ \\
$\rm \sigma_{X-ray}$ & 6.61 $\rm \times 10^{-25}\ [m^2]$ \\
Speed of light & 299 792 458  $\rm [m.s^{-1}]$ \\

\hline
\multicolumn{2}{c}{Simulation specs} \\
\hline
Comoving resolution dx & 1 $[h^{-1}.\rm{cMpc}]$ \\
DM particle mass &  1.075 $\rm \times 10^{11} [M_{\odot}]$  \\

\end{tabularx}
\caption{ \label{Tab:specs} \textbf{Specifications of simulation:}  The cosmological parameters are extracted from \citep{Planck} Tab. 2 last column (and $\rm{\Omega_b} = \rm{\Omega_b}h^2 / h^2$).  }
\end{table}

\subsubsection{Results}
\label{subSec:Results}

\begin{figure*}
\includegraphics[width=1.\columnwidth]{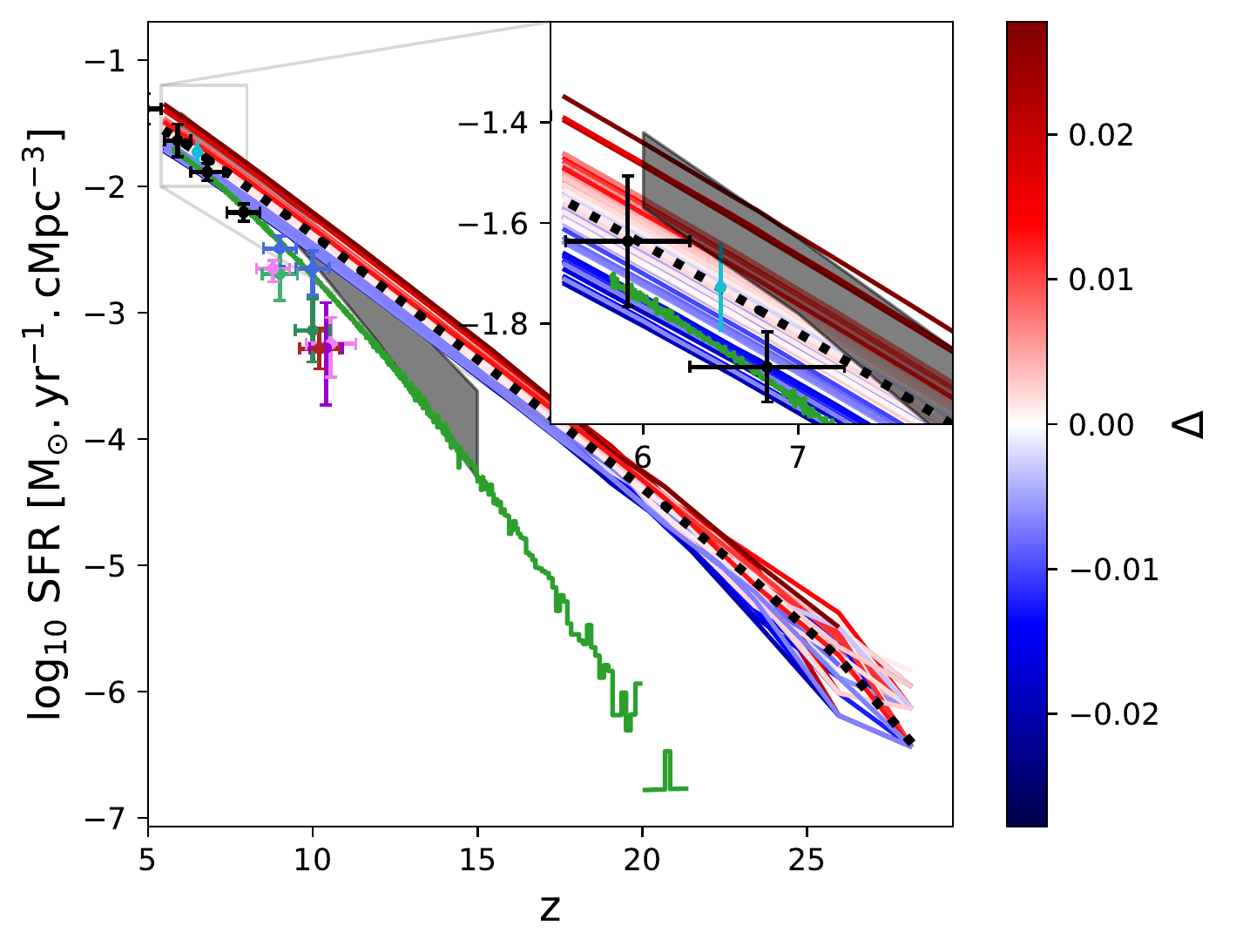} 
\includegraphics[width=1.\columnwidth]{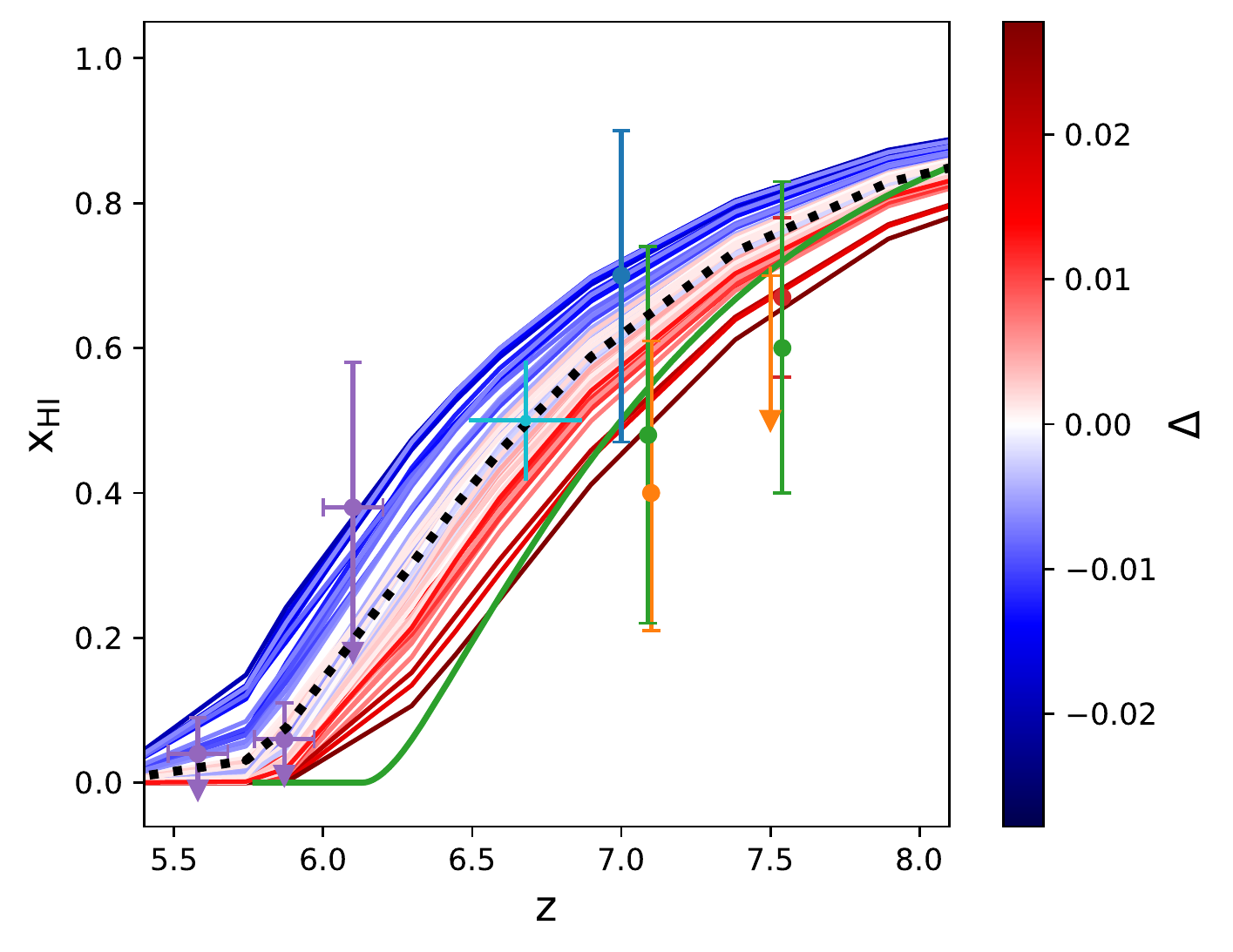} 
\caption{ \textbf{SFR and neutral fraction}: on the left panel the evolution with redshift of the cosmic star formation rate.  The average is shown with the doted black thick line.  The blue and red lines presents the cSFR for sub-cubic volume of $64h^{-1}.\rm{cMpc}$ side, the color code for under and over dense region, respectively.  The cyan error bar in the inset illustrate the 1-sigma dispersion induce by the large scale density fluctuations at redshift 6.5.  The observation points comes from different probes.  On the left panel the constrains on the cosmic SFRD are: \citealt{Bouwens2014,Bouwens2016} in black and violet, \citealt{McLeod2016} in blue, \citealt{Oesch2013,Oesch2014,Oesch18} in green and brown and \citealt{Ishigaki18} in pink.  On the right panel the constrains on the neutral fraction are: \citealt{McGreer2015} are in purple, \citealt{Greig2017,Greig2019} are in orange, \citealt{Davies2018} are in green, \citealt{Banados2018} is in red and \citealt{Wang2020} in dark blue.  }
\label{Fig:dispertionSFR}
\end{figure*}

The cosmic SFR is calibrated to be roughly on or above of the observations at redshift 6 ($ 3\times 10^{-2} \rm{ [M_{\odot}.yr^{-1}.cMpc^{-3}]  }$) and $ 10^{-6} \rm{ [M_{\odot}.yr^{-1}.cMpc^{-3}]  }$ at redshift 30.  It accounts for the fact that the cosmic SFR predicted by the simulation contains the contribution of all the galaxies, while the observations are limited to magnitude -17.  Fig. \ref{Fig:dispertionSFR} presents the cosmic SFR on the left panel and the neutral fraction on the right.  The gray area presents the estimated total SFR \citep{Gillet2020}.  In the simulation, the evolution of the cosmic SFR with redshift is induced by the evolution of the density distribution and the evolution of the parameter $\Delta_{\Sigma_*}$. The ionization history is calibrated in order to have a mid reionization between redshift 6 and 7.  The CoDaII averages are also shown in green for comparison.  Even with its mass/spatial resolution, the CoDaII is not able to from stars at the early redshift (z=30).  Here, the new parametrization is able to form stars at the CD, while encompass the sub-grid feedback on SFR at later redshift.

Additionally, Fig. \ref{Fig:dispertionSFR} presents the dispersion of the SFR and neutral fraction for sub-cubic-volumes of 64\hcMpc side that can be compared to the volume of the CodaII simulation that was used to calibrate the star formation model.  The over density of each sub-volume is indicated in red and blue for over and under-dense region respectively.  The dispersion in SFR is relatively constant between z=6 and 30, and is comparable to the observations uncertainties (illustrated at redshift 6.5 with the cyan error-bars).  In the case of the neutral fraction, the dispersion at mid-reionization is slightly smaller that current observations estimation with $\pm 0.08$ and the redshift dispersion is about $\pm 0.19$ around the average mid-ionization redshift (illustrated with the cyan error-bars). Overall, these results demonstrate that our new star formation model can be made consistent with constraints during the EoR, while providing a sustained star formation during the Cosmic Dawn.


\section{21cm signal}
\label{Sec:21cm signal}


Additionally to the new star formation prescription, we added in the code the computation of the 21cm signal. The goal is to predict the possible 21cm signal that could observe radio telescopes from the Cosmic Dawn to the end of the Reionization.  For those kind of observations, high resolution are not needed, 1\hcMpc of resolution is enough. But a large volume is require to probe the largest mode that will be observed.  The following results are presented for the largest box available in this study: 512\hcMpc.  


\subsection{Simulation of the signal}
\label{Sec:Simulation of the signal}


The formula of the 21cm brightness temperature with respect to the CMB at a given redshift and point in space is given by:

\begin{equation}
\begin{split}
    \rm{\delta} T_{21} & \approx 27(1-x_{\rm{HII}}) (1+\delta) (1-\frac{T_{\rm{CMB}}(z)}{T_{\rm{s}}}) C_{\rm{cosmo}} \  [\rm{mK}] \\
    C_{\rm{cosmo}} & = \left( \frac{\rm{\Omega_b}}{0.044} \right) \left( \frac{h}{0.7} \right) \sqrt{ \frac{0.27}{\rm{\Omega_m}} } \sqrt{ \frac{1+z}{10} }
\end{split}
\label{Eq:T21}
\end{equation}

where $x_{\rm{HII}}$ is the ionized fraction of the gas, $\delta$ its over-density, $T_{\rm{CMB}}$ the temperature of the CMB and $T_{\rm{s}}$ the spin temperature.  We neglect the velocity gradient in this study. The spin temperature of the gas can be computed from:

\begin{equation}
\begin{split}
    T_{\rm{s}} & = \frac{1 + x_{\rm{c}} + x_{\rm{\alpha}} }{ T_{\rm{CMB}}^{-1} + x_{\rm{c}} T_{\rm{K}}^{-1} + x_{\rm{\alpha}} T_{\rm{c}}^{-1} }
\end{split}
\label{Eq:Ts}
\end{equation}

where $T_{\rm{K}}$ is the kinetic temperature of the gas, $T_{\rm{c}}$ the color temperature of the radiation field at the \Lya transition, $x_{\rm{c}}$ is the collision coupling coefficient and $x_{\rm{\alpha}}$ is the coupling coefficient associated with \Lya pumping.  

In this study we do not include the \Lya radiative transfer, therefore in the following we will consider two regimes. At first we consider a uniform \Lya coupling factor rising with redshift due to a rising LyA background: (${\rm{log_{10}}}(x_{\rm{\alpha}})=-3/8\ z+7.25$) which mimic the average evolution from Fig 2 of \cite{Ross2019}.  By doing so we can produce realistic global temperature evolution, but the power spectrum cannot take into account the spatial fluctuations of $x_{\rm{\alpha}}$.  We also consider the saturated regime, where we assume $x_{\rm{\alpha}} \gg 1 + x_{\rm{c}}$ everywhere and $T_{\rm{s}} = T_{\rm{c}} = T_{\rm{K}}$.

Finally, the collision coupling coefficient accounts for the H-H, H-$\rm{e^-}$ and H-$\rm{H^+}$ collisions and is given by:

\begin{equation}
\begin{split}
    x_{\rm{c}} = \frac{\rm{T_{10}}}{\rm{A_{10}}} \frac{1}{T_{\rm{CMB}}(z)} (n_{\rm{HI}} \kappa_{\rm{HH}}+n_{\rm{p}}\kappa_{\rm{pH}}+n_{\rm{e}}\kappa_{\rm{eH}}),
\end{split}
\label{Eq:xc}
\end{equation}

$\kappa_{\rm{i}}$ are the spin de-excitation rates for each type of collisions and $n_{\rm{i}}$ the densities, $\rm{T_{10}}=0.068$[K] and $\rm{A_{10}}=2.85\times10^{-15} \rm{[s^{-1}]}$ is the spontaneous emission rate.  The de-excitation rates are taken into account as follow:

\begin{itemize}

    \item $\kappa_{\rm{HH}}$ is interpolated from \cite{Zygelman2005} Table 2 column 4 for $1\rm{K}\leq T_{\rm{K}}\leq 300\rm{K}$ or $\kappa_{\rm{HH}}=3.1\times 10^{-11}T_{\rm{K}}^{0.357}e^{-32/T_{\rm{K}}}\ \rm{[cm^{3}s^{-1}]}$ for $300\rm{K} \leq T_{\rm{K}}$ \citep{Kuhlen2006}.
    
    \item $\kappa_{\rm{eH}}$ is interpolated from \cite{Furlanetto2007a} Table 1 for $1\rm{K}\leq T_{\rm{K}}\leq 10000\rm{K}$ or $\rm{log_{10}}(\kappa_{\rm{eH}})\approx -8.0958$ for $10000\rm{K} \leq T_{\rm{K}}$ \citep{Liszt2001}.
    
    \item $\kappa_{\rm{pH}}$ is interpolated from \cite{Furlanetto2007b} Table 1 for $1\rm{K}\leq T_{\rm{K}}\leq 20000\rm{K}$ or $\kappa_{\rm{pH}}= 2\kappa_{\rm{HH}}$ for $20000\rm{K} \leq T_{\rm{K}}$.
    
\end{itemize}

The 21cm signal is computed on the fly by the EMMA simulation code for the two \Lya regimes (saturated and average background). The power spectrum (PS) of the simulated temperature brightness fields are computed using \code{tools21cm} \citep{tools21cm} in post-processing. The spherically average dimensionless power spectrum ($\Delta^2(k)$) is computed using: 
\begin{equation}
\Delta^2(k) = \frac{k^3}{2\pi^2} \langle P(\textit{\textbf{k}}) \rangle_{(k_x,k_y,k_z)},
\label{Eq:PS}
\end{equation}

where $P(\textit{\textbf{k}})$ is the power spectrum, and $k_i$ are the components of the wave-vector along the simulation volume.


\subsection{Observation of the signal}
\label{Sec:Observation of the signal}


Being in possession of an 'ideal' noiseless 21cm PS from the cosmic dawn, we used \href{https://gitlab.com/flomertens/ps_eor}{\code{ps\_eor}} \footnote{\url{https://gitlab.com/flomertens/ps_eor}} to take into account the UV coverage and the noise level due to the instrument. 
We focus on the New Extension in Nançay Upgrading loFAR (NENUFAR; \cite{NENUFAR2012}) observations as we are part of the NENUFAR Cosmic Dawn key project.  NENUFAR is a radio interferometer that will observe between 85Mhz and 30MHz, covering the CD epoch.  Interestingly it covers the 83-73 MHz band where the EDGES collaboration reported a signal detection \citep{Bowman2018}.   

Radio interferometers may produce 3D data-cube, 2D on the sky and the third dimension corresponding to the frequency that can be converted in distance/redshift/time assuming a cosmological model.  To get as close as possible to the observations we have to construct a data-cube corresponding to the same coverage on the sky and depth in frequency.  The observations specifications are listed in Tab. \ref{Tab:Obs_specs} and correspond to the ongoing Cosmic Dawn observation program made with NenuFAR.  We focus on the highest frequency band, centered on redshift 17 (corresponding to the EDGES's claimed detection band).  The shape of the observed volume is 2982.29 cMpc on the sky direction and 231.54 cMpc in depth.  The volume is divided in $68^2$ pixels on the sky and 51 along the line of sight.  As the depth of the data-cube is relatively small (231.54 cMpc) we neglect for the moment the increase of the size with the depth, as well as the time evolution along the frequency (light-cone effects) \citep{Greig2018}: the simulation size (756 cMpc) is larger than the observational depth, a third of box is enough in depth.  Conversely, on the sky's axes, the box is repeated $\sim 4$ times.  It should be noted that the observed modes are overwhelmingly due to $k_{\parallel}$, which correspond to the line of sight.  The $k_{\parallel}$ modes are roughly 1 order of magnitude greater than $k_{\perp}$.  Therefore the result is not affected by the periodic repetition of the box.  Once the mock data-cube is filled by the simulation it is given to \code{ps\_eor} to compute the PS and the theoretical thermal noise level.

\begin{table}
\begin{tabularx}{\columnwidth}{ 
>{\raggedright\arraybackslash} X 
>{\centering\arraybackslash} X}
 \hline
 \multicolumn{2}{c}{NENUFAR observations specs} \\
 \hline
 Band-width           & 9,96 [MHz] \\
 channel-width        & 195.3 [kHz] \\
 Number of channel    & 51 \\
 redshift at center   & 17 \\
 frequency at center  & 78.91 [MHz] \\
 BW limits            & 83.79-73.83 [MHz] \\
 BW limits redshift   & 15.95-18.24 \\
 Depth                & 231.54 [cMpc] \\
 Depth resolution     & 4.54 [cMpc] \\
 \\
 Field of view        & 16 [$\degree$] \\
 FoV at center        & 2982.29 [cMpc] \\
 Number of pixels across the sky & $68^2$ \\
 Sky resolution       & 43.857 [cMpc] \\
 \\
 Total obs time       & 1000 [h] \\
 Time obs per day     & 8 [h] \\
 Integration time     & 100 [s] \\
 
\end{tabularx}
\caption{ \label{Tab:Obs_specs} \textbf{Observation specifications:} first the frequency, secondly the sky and finally the observation time information. The transformations to comoving distance are made at the central redshift, the data-cube is consider as 'cubic'. }
\end{table}


\subsection{Simulated observations of the 21cm}
\label{Sec:Simulated observations of the 21cm}


\begin{figure*}
\includegraphics[width=.32\textwidth]{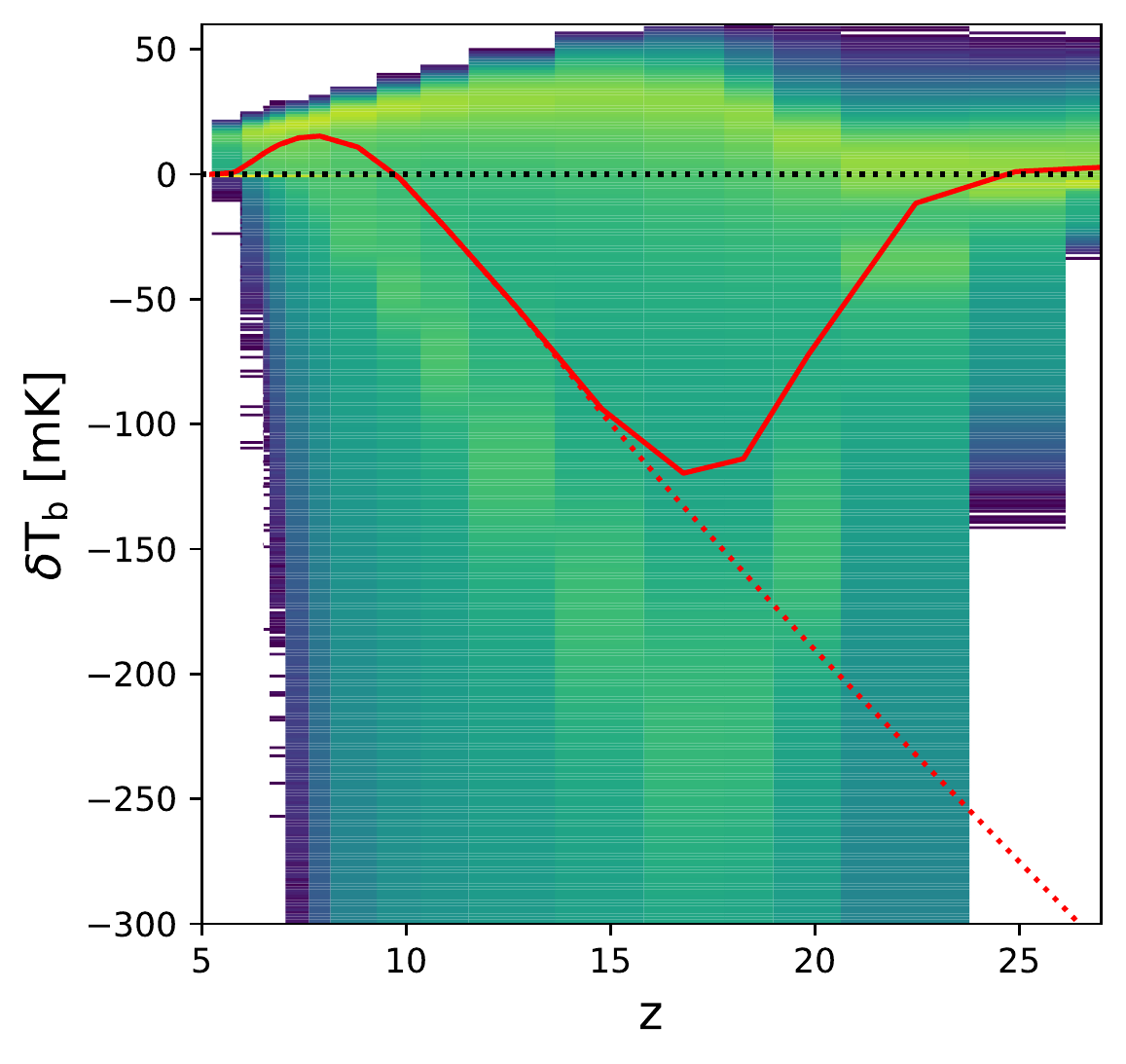}
\includegraphics[width=.37\textwidth]{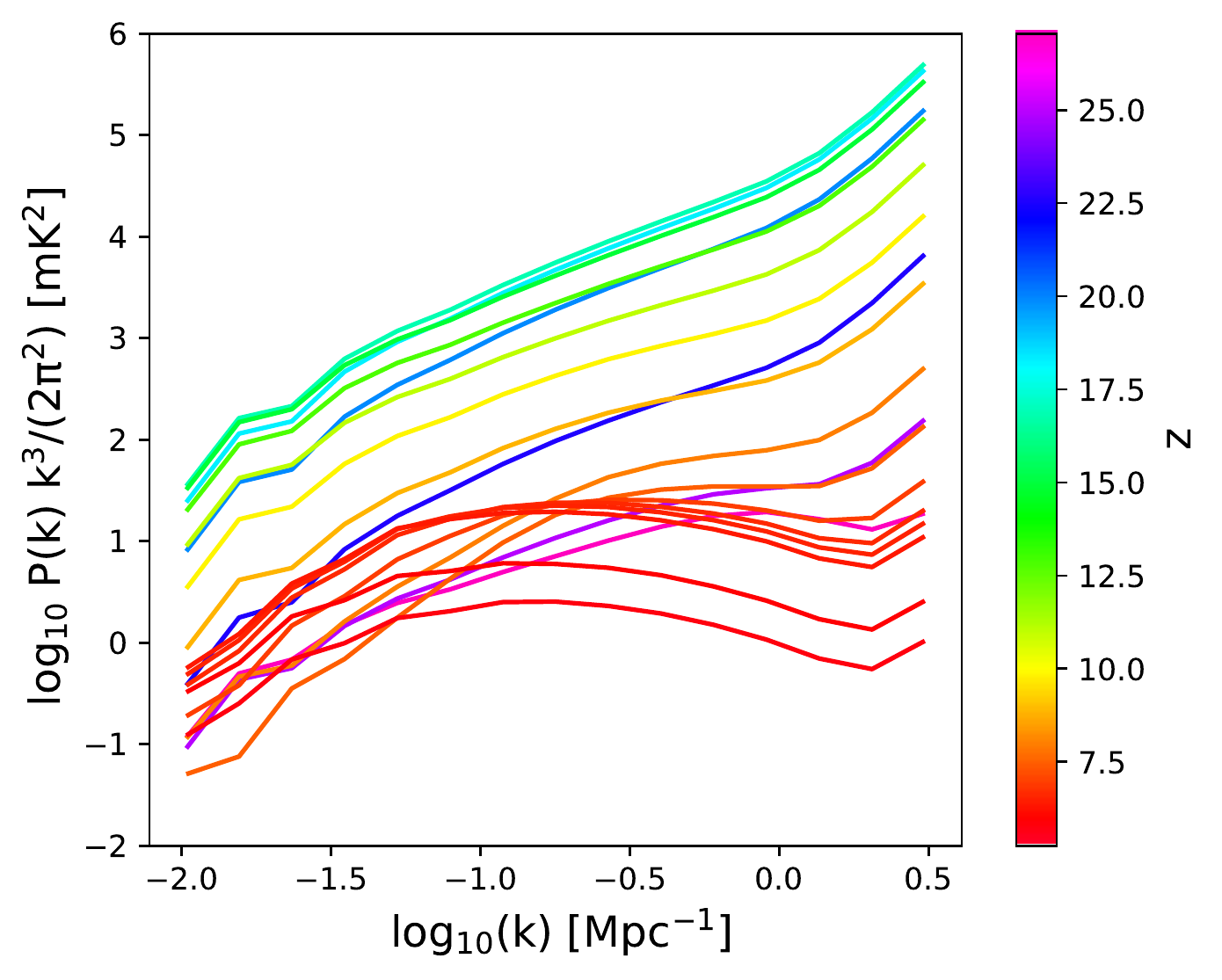}
\includegraphics[width=.30\textwidth]{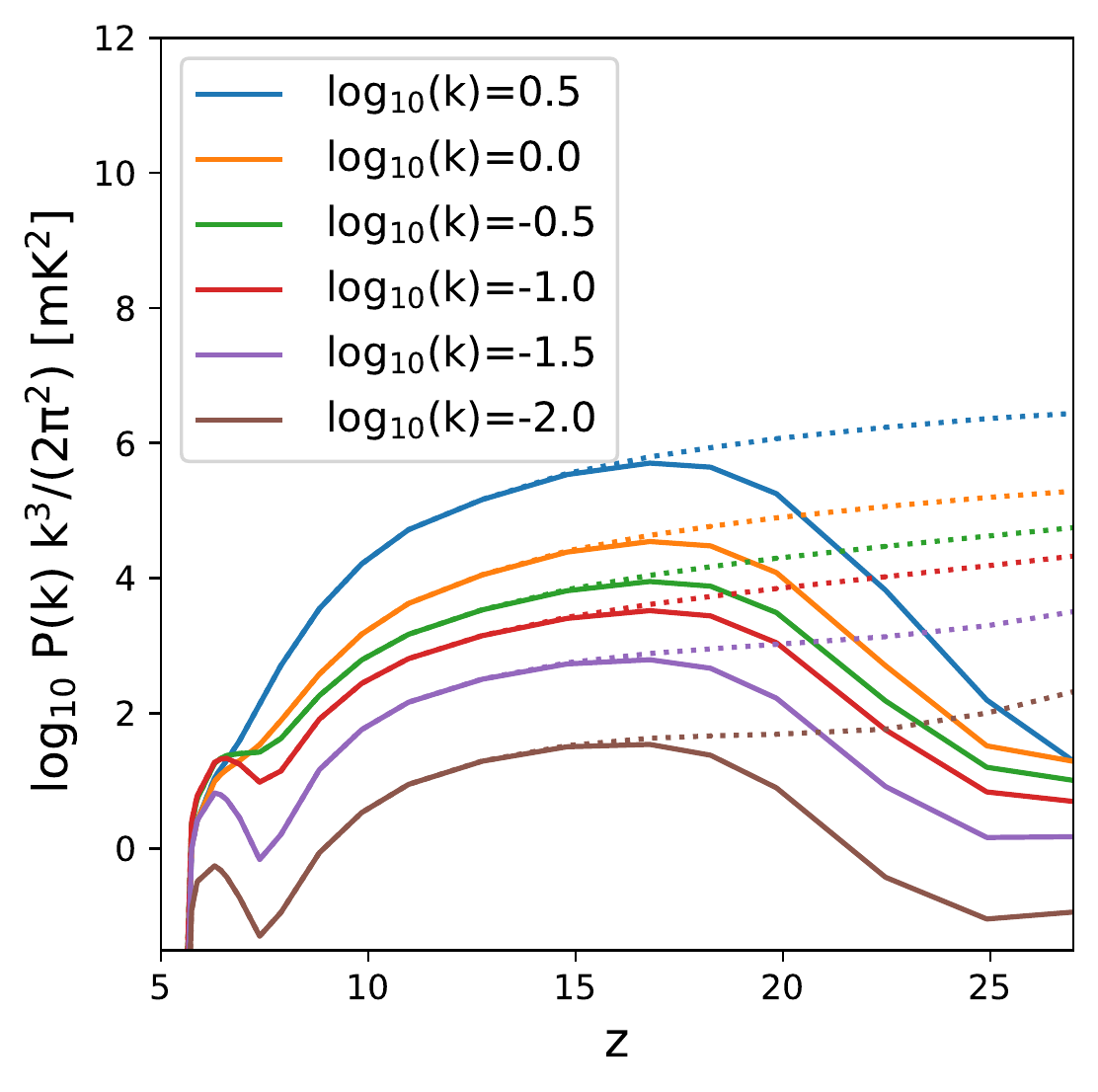}
\caption{ \textbf{The 21cm signal}: On the left panel, the distribution of the 21cm temperature brightness as function of redshift. 
The red full line presents the average evolution of the brightness temperature taking into account of the uniform \Lya{} background, while the red dotted line assume a fully coupled gas and spin temperature.  The background color code for the volume weighted distribution.  On the middle panel, the power spectrum at all scale at different redshifts (the color code redshift between 6 to 30) in the case of the \Lya uniform background.  While, the right panel presents the power evolution with redshift of some specific scales ($\rm{log_{10}k=}0.5,0,-0.5,-1,-1.5$), for the \Lya uniform background and fully coupled approximation in full and dotted lines respectively. }
\label{Fig:PS}
\end{figure*}

After the calibration of the SFR and ionization history (c.f. \ref{subSec:Results}) we analyze the 21cm signal.  Fig. \ref{Fig:PS} presents different quantities related to the 21cm signal.  On the left panel, the global average brightness temperature is shown in red.  The background color shows the distribution of the brightness temperature with redshift (volume weighted).  We note that the brightness temperature is bi-modal between redshift 21 and 8, with a cold and a hot phase.  The middle panel presents the power spectrum (for coeval cubes, i.e. not taking into account light-cone effects) at different redshifts and the right panel presents the evolution of some specific $\Delta^2(k)$ with redshift.  The PS presented here are qualitatively similar to simulated expectations (see \citealt{Greig2017, Ross2019, Itamar2020} for examples).  Note that above redshift 15 the PS is affected by the missing Lyman-$\alpha$ transfer.   The uniform \Lya back-ground reduce uniformly the power at every scale above redshift 15, illustrated on the Fig. \ref{Fig:PS} right panel with the full lines and the dotted lines illustrate a full \Lya coupling at all time.  While, the propagation of the \Lya photons thought the IGM should induce spatial patterns and so different power evolution with redshift.

Finally, the main goal is to produce a 21cm PS as close as possible to the future observed one.  We process this cube through \code{ps\_eor} in order to take into account of the UV coverage (see Sec. \ref{Sec:Observation of the signal}).  In theory, the PS outputted by \code{ps\_eor} should be the same as the one obtain on the 'perfect' simulated cubes, in the range of scale well sampled, and in the absence of further distortion.  In the present study we do not include other source of noise subtraction or distortion on the signal, like wedge treatment or foreground residuals.  The wedge is a portion of the Fourier space where the foreground signal due to the galaxy is dominant.  There are two main strategies to extract the 21cm cosmic signal.  The first, the wedge avoidance, consist to cutoff the data where the galactic foreground is too dominant.  The resulting PS estimation should be foreground free, but some peaces of the signal are lost as some data have been deleted.  The second, the foreground modeling, consist to try to keep all the data by modeling the foreground and substrate it.  It as the advantage to conserve more data, therefore more signal, but at the cost of some modeling dependencies and foreground residuals which are difficult to quantify.

\begin{figure}
\includegraphics[width=\columnwidth]{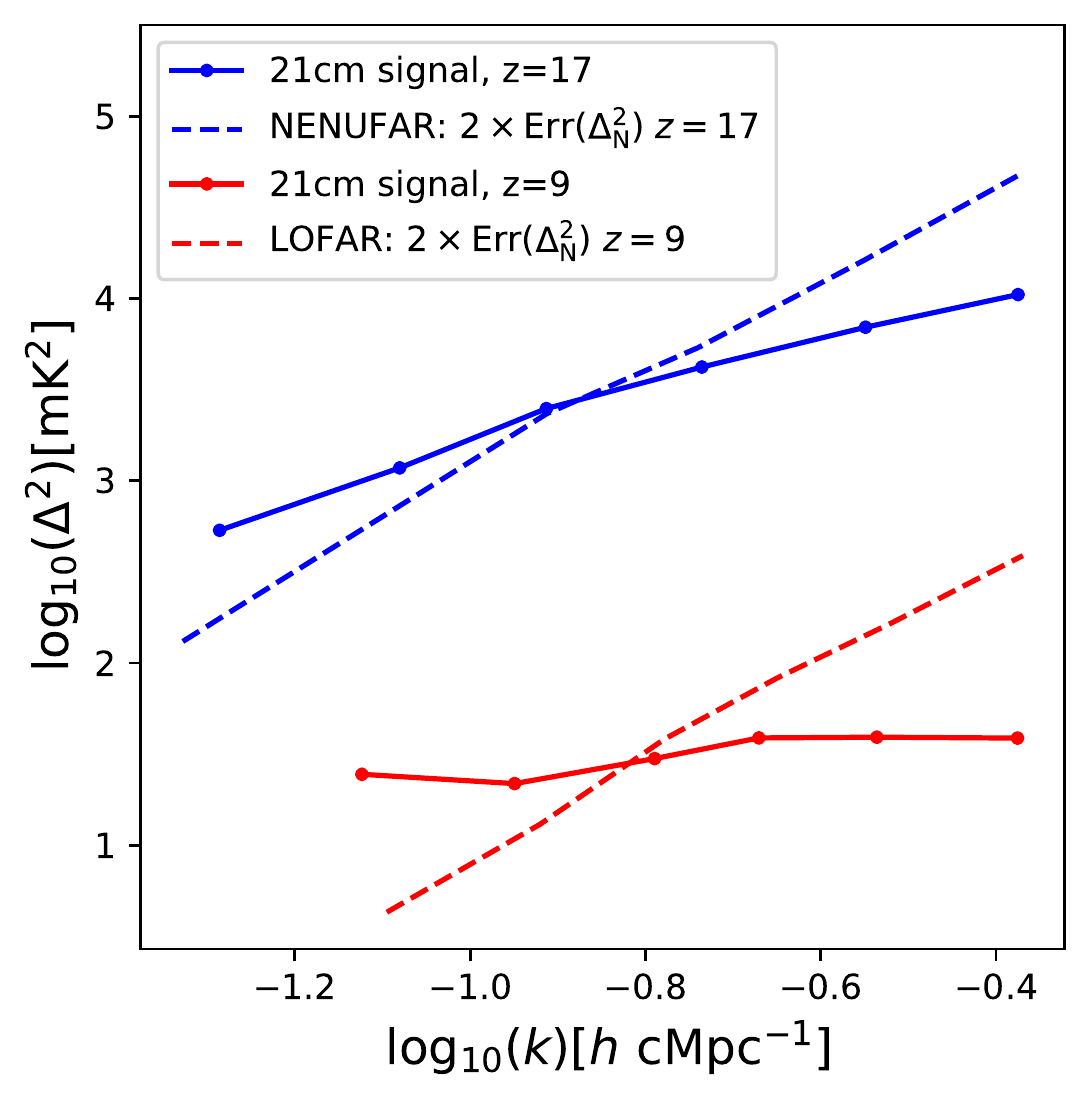}
\caption{ \textbf{The 21cm power spectrum}: the dotted full lines present the power spectrum given by \code{PS\_EOR}, which take into account of the resolution and the UV coverage of the instruments, NENUFAR and LOFAR, respectively in blue and red at redshift 17 and 9.  The dashed lines present the expected 2$-sigma$ sensitivity  for 1000h of observations. }
\label{Fig:PS_obs}
\end{figure}

On Fig. \ref{Fig:PS_obs} we present in blue the PS at redshift 17 and the $2-\sigma$ theoretical error due to the thermal noise (dashed blue line).  We also present the predicted PS  at redshift 9 and the error for the LOFAR.  In both cases, a detection is expected for wavenumber below $k=0.1\ {\rm h\ cMpc^{-1}}$.  The most recent upper limit at redshift 9 of $\rm{log_{10}(\Delta^2)=3.73}$ at $k=0.075\ {\rm h\ cMpc^{-1}}$ \citep{Mertens2020} is 2 dex above our prediction and at redshift 17 $\rm{log_{10}(\Delta^2)=782}$ at  $k=0.15\ {\rm h\ cMpc^{-1}}$ \citep{Gehlot2020} is 4 dex above our prediction (not added on Fig. \ref{Fig:PS_obs}). 

\section{Conclusions}
\label{Sec:Conclusions}


In this paper we introduce a new large scale galaxy formation model in the fully coupled dark matter, hydrodynamics, radiative transfer code EMMA.  This empirical model allows the efficient production of large scale low resolution simulations of the CD and EoR with a reduce and flexible set of parameters, based on the results of the state of the art simulation of the Reionization CoDaII. 
We ran a simulation using this model and predict the associated 21cm signal.  We process it up to the prediction of the power spectrum with tools as close as possible to the one used to reduce the observational data.  The resulting power spectrum obtained on a $(512 h^{-1}\rm{cMpc})^3/512^3$ elements of resolution fiducial simulation are qualitatively comparable to state of the art predictions.  

We focused on the ongoing observations of the radio telescope NENUFAR, that is covering the cosmic dawn.  We predict that our fiducial model should be detected by NENUFAR at redshift 17 at wavenumber between $k=0.1\ {\rm h\ cMpc^{-1}}$ and $k=0.06\ {\rm h\ cMpc^{-1}}$ with 1000h of observations.  LOFAR should detect the signal at the same wavenumber at redshift 9. 

While waiting for the data acquisition, reduction and analysis we plan to explore the parameter space.  Specifically, the next step is to quantify how much a signal detection at $k=0.1\ {\rm h\ cMpc^{-1}}$ and $k=0.06\ {\rm h\ cMpc^{-1}}$ at redshift 17 may constrain our parameters, for example the SFR spatial distribution.  A large number of points still have to be addressed, such as, the inclusion of \Lya photons is essential for the computation of the 21cm signal, or a sub-grid treatment of the temperature to take into account of the sub-cell multi-phase of the gas \citep{Ross2019}. 

\section*{Acknowledgements}
We thank Anastasia Fialkov for fruitful discussions and sharing data to help the validation of the model.  We thank the CoDa Collaboration for sharing the data of the CoDaII simulation.  

NG is supported by the University of Strasbourg IDEX post-doctoral grant “Predicting with cosmological simulations the 21cm signal from the Epoch of Reionization for future large radio observatories”.

This work was granted access to the HPC resources of CINES under the allocations 2020-A0070411049 and 2021- A0090411049 “Simulation des signaux et processus de l’aube cosmique et Réionisation de l’Univers” made by GENCI.

This research made use of \code{astropy}, a community-developed core Python package for astronomy \citep{Astropy}; \code{matplotlib}, a
Python library for publication quality graphics \citep{Matplotlib}; \code{scipy}, a Pythonbased ecosystem of open-source software for mathematics, science, and engineering \citep{SciPy-NMeth} – \code{numpy} \citep{NumPy-Array} and \code{Ipython} \citep{ipython} 

\bibliographystyle{mnras}
\bibliography{bib}




    
    
    


\bsp	
\label{lastpage}
\end{document}